\documentclass[prd, twocolumn, showpacs]{revtex4-1}

\usepackage{graphicx,amsmath,amssymb}
\usepackage{dcolumn}
\usepackage{bm}

\begin{document}

\title{Strong gravitational lensing by Schwarzschild black hole}

\author{Tushar kanti Dey}
\email[]{tkdey54@gmail.com}

\affiliation{Physics Department, Guru Charan College, Silchar - 788004, India}


\begin{abstract}
Schwarzschild black holes can produce strong gravitational lensing. The relativistic images are produced due to bending of light around the black hole.  We propose a model equation to study the strong gravitational lensing. The model equation can well describe the bending angle from a range very close to the photon sphere to the first relativistic image.

\end{abstract}

\pacs{95.30.Sf, 04.70.Bw, 98.62.Sb}


\maketitle

\section{\label{1x} Introduction}

It is a challenging  job to compute the bending angle in the strong gravitational field. Since the work of Darwin ~\cite{Darwin59, Darwin61}  there were various attempts to find the deflection angle in the strong field limit. Also recent works by V. Bozza, S. Capozziello, G. Iovane and G. Scarpetta ~\cite{Bozza2001}, V. Bozza ~\cite{Bozza2002, Bozza2010}, P. Amore and S. Arceo ~\cite{Amore2006}, P. Amore and M. Cervantes ~\cite{Amore2007}, K. S. Virbhadra and G. F. R. Ellis ~\cite{Virbhadra2000},  S. V. Iyer and A. O. Petters ~\cite{Iyer2007} made considerable progress in calculating the deflection angle in strong gravitational field. Relativistic images produced by strong gravitational lensing (SGL) will provide a good observational test for the general theory of relativity in the strong field. It is believed that relativistic images produced by black holes, which is $ \sim\mu as$, will be within our observational reach in near future. Very Long Baseline Interferometry (VLBI) ~\cite{VLBI, Ulvestad1999} may resolve relativistic images. An analytical expression for the deflection angle will be very helpful in describing the sizes of the relativistic images. In this paper we have proposed a model equation for a Schwarzschild object which fit exact deflection angle with minor error. The range of validity is very near the photon sphere to the first relativistic image so that relativistic images can be investigated properly.

\section{\label{2x} The deflection angle}

The exterior gravitational field of an object with spherical symmetry is given by the Schwarzschild line element
{\setlength\arraycolsep{2pt}
\begin{eqnarray}\label{eq:one}
  ds^2 &=& \left(1-\frac{2M}{r}\right)dt^2 - \left(1-\frac{2M}{r}\right)^{-1} \nonumber \\
   & & - r^2\left(d\theta^2 + sin^2\theta d\phi^2\right)
\end{eqnarray}}

Here we consider $G=c=1$ and $M$ is the enclosed mass. The Schwarzschild radius is $R_{sch}=2M$.
The bending angle $\hat{\alpha}$ with closest distance of approach $r_0$ is given by  ~\cite{Weinberg}
\begin{equation}\label{eq:two}
\widehat\alpha (r_0 ) = 2\int_{r_0 }^\infty  {\frac{{dr}}{{r\sqrt {\left( {\frac{r}{{r_0 }}} \right)^2 \left( {1 - \frac{{2M}}{r_0}} \right) - \left( {1 - \frac{{2M}}{r}} \right)} }}}  - \pi
\end{equation}
and the impact parameter $J$ of the light ray is given by
\begin{equation}\label{eq:three}
    J(r_0 ) = \frac{r_0 }{\sqrt {1 - \frac{2M}{r_0 }} }
\end{equation}
Now we define the dimensionless parameter $x$ and $x_0$ by
\begin{equation}\label{eq:four}
x = \frac{r}{{2M}} \; , \qquad  x_0 = \frac{r_0}{{2M}}.
\end{equation}
The deflection angle (Eq.~\eqref{eq:two}) and impact parameter (Eq.~\eqref{eq:three}) become
\begin{equation}\label{eq:five}
\widehat\alpha (x_0 ) = 2\int_{r_0 }^\infty  {\frac{{dx}}{{r\sqrt {\left( {\frac{x}{{x_0 }}} \right)^2 \left( {1 - \frac{1}{{x_0 }}} \right) - \left( {1 - \frac{1}{x}} \right)} }}}  - \pi
\end{equation}
and 
\begin{equation}\label{eq:six}
J(x_0 ) = \frac{{2Mx_0 }}{{\sqrt {1 - \frac{1}{{x_0 }}} }}
\end{equation}
The deflection angle  is an elliptical integral. So numerical integration is the only method to find the exact deflection angle. But an analytical expression is very handy to get the deflection angle. We propose the following equation which gives deflection angle very accurately near the photon sphere.
\begin{equation}\label{eq:seven}
\hat \alpha_{_D} \left( {x_0 } \right) = {\rm{a }}\; \text{ ArcCsch}({\rm{b }}\;  x_0  +  {\rm{c }}){\rm{ }} + {\rm{ d}}
\end{equation}
where\\
\begin{align*}
a& =-1.9904142396804585&
b& =-7.152403979116706\\
c& =\quad 10.728609274134325&
d& =\quad 2.603634899810115
\end{align*}
The constants $a$, $b$, $c$ and $d$ are calculated using the software Mathematica. 
For Schwarzschild lens the photon sphere is at $x_{\rm{ph }}=1.5$. The range of validity of this equation is that the value of  $x_0$ is from very near the photon sphere to  $1.55$ . Eq.~\eqref{eq:two} is an elliptic integral, but  Eq.~\eqref{eq:seven} gives an approximate solution to this integral. Now we shall calculate the impact parameter and distance of closest approach  for relativistic Einstein rings which correspond to deflection angles, $\hat \alpha_{_D}$,  equal to $2 \pi, \; 4 \pi, \; 6 \pi, \; \cdots$.
{\setlength\arraycolsep{2pt}
\begin{eqnarray}\label{eq:eight}
\frac{J}{M}- 3\sqrt{3} &=&0.00653693496, \,  0.00001213266,\nonumber \\ & & 2.22951008765 \times 10^{-8}, \nonumber
  \\ & &  5.1546322765\times 10^{-11}, \nonumber \\ & & 1.27542421068\times 10^{-12},  \cdots
\end{eqnarray}}
{\setlength\arraycolsep{2pt}
\begin{eqnarray}\label{eq:nine}
\frac{r_0}{M}- 3 &=& 0.09029475975, \, 0.00374917067,\nonumber \\ & &  0.00016046132, \, 7.71499009477\times10^{-6}, \nonumber \\ & & 1.21333938496\times10^{-6}, \cdots .
\end{eqnarray}}
\noindent
We have compared the values of Eqs.~\eqref{eq:eight},~\eqref{eq:nine} with that given by Ref.  ~\cite{Bisnovatya, Misner}.
The comparison of different values of the impact parameter with the values given in references reveal that the the model Eq.~\eqref{eq:seven} can well describe the strong field limit.

\begin{figure}
\includegraphics[width=7cm, height=.2\textheight]{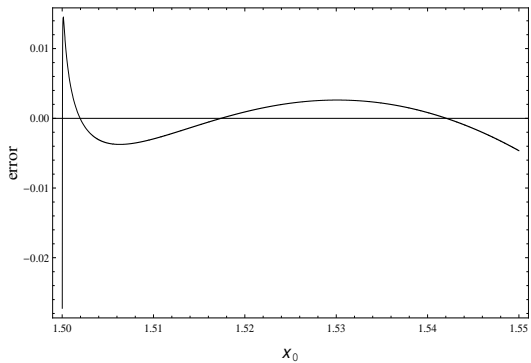}
\caption{\label{fig:error} The error in deflection}
\end{figure}

\begin{figure}
\includegraphics[width=7cm, height=.2\textheight]{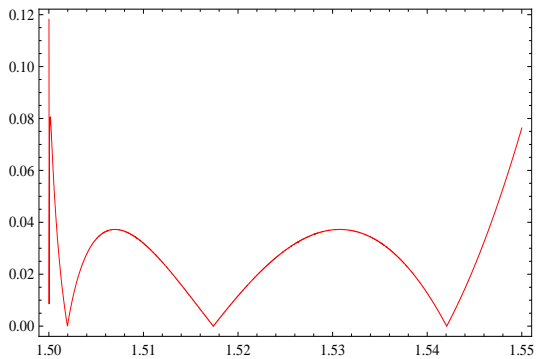}
\caption{\label{fig:errorpercentage} The percentage of error in deflection angle}
\end{figure}

The exact value of the deflection angle corresponds to the numerical integration of the deflection angle (Eq.~\eqref{eq:two}).The error in the deflection angle defined as $(\hat{\alpha_D} - \hat{\alpha})$ is shown in Fig. ~\ref{fig:error}. Fig. \ref{fig:errorpercentage} shows the percentage of error defined as $100\times |\frac{\hat{\alpha_D} - \hat{\alpha}}{\hat{\alpha}}|$. The table ~\ref{tab:table1} also show the error of $\hat\alpha_{_D}$ and percentage of error for different value  $x_0$ between $1.50001$ to $1.55$.

\begin{table}
\caption {\label{tab:table1}
 The error defined as $(\hat{\alpha_D} - \hat{\alpha})$ and percentage of error defined as $100\times |\frac{\hat{\alpha_D} - \hat{\alpha}}{\hat{\alpha}}|$ in deflection angle for different values of $x_0$ }
\begin{ruledtabular}
\begin{tabular}{lcc}
$x_0$ &
\textrm{error}  &
\textrm{percentage of error}\\
\colrule
1.50001 & -0.0272447 & 0.01182\\
1.5001 & 0.0143755 & 0.07803\\
1.5010 & 0.0041091 & 0.02973\\
1.50196 & $0$ & 0\\
1.51   & -0.0029485  & 0.03193\\
1.52   & 0.0009176 & 0.01166 \\
1.53   & 0.0026232 & 0.03710\\
1.54   & 0.0008290 & 0.01272\\
1.55   & -0.0046430 & 0.07628
\end{tabular}
\end{ruledtabular}
\end{table}

We also draw graphs of the deflection angle as proposed by different authors ~\cite{Bozza2010, Amore2007} with that of our model equation to show how our equation describe the deflection angle in the strong field limit. In the following we shall write down the expression of deflection angle as given by Bozza and Amore.

\begin{figure}
\includegraphics[width=7cm, height=.3\textheight]{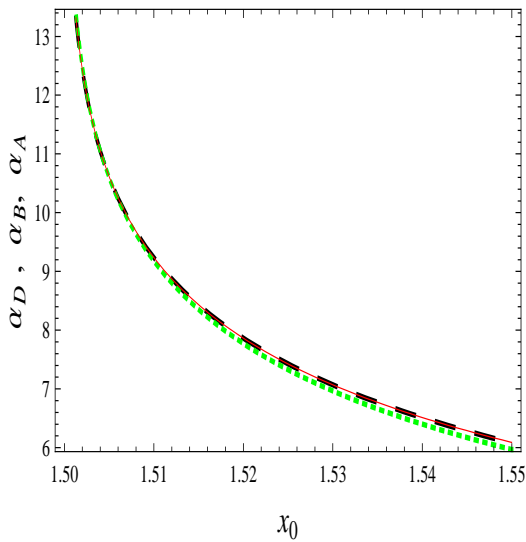}
\caption{\label{fig:comparison} The comparison among the deflection angles due to Bozza $ \, \alpha_B$  , Amore \textit{et al} $ \, \alpha_A$  and model equation $ \, \alpha_D$}
\end{figure}

The deflection angle due to a Schwarzschild object in the strong deflection limit as proposed by Amore ~\cite{Amore2007} is
\begin{equation}\label{eq:amore}
    \alpha_{A}=\frac{12}{-3\sqrt{4\zeta-1}-4}+\sqrt{4\zeta+\frac{1}{3}}\log\left(\frac{\zeta}{\zeta-1}\right)
\end{equation}
and that due to Bozza ~\cite{Bozza2010} is
\begin{equation}\label{eq:bozza}
    \alpha_{B}=-\log\left(\frac{J}{\overline{J}}-1\right)+\log[216(7-4\sqrt{3}\, ]-\pi
\end{equation}
\noindent
where $\overline{J}$ is the smallest impact parameter and its value is $ 3 \sqrt{3}M/2 $ and $\zeta=2 x_0/3$.
Fig \ref{fig:comparison} shows the comparison of the curves drawn using Eq.~\eqref{eq:amore} (dotted green curve), Eq.~\eqref{eq:bozza} (thick dashed black curve) and the model  Eq.~\eqref{eq:seven} (continuous red curve). It is seen that the curve of the deflection angle due to Bozza is the same as due the model equation but a little difference with the curve due to Amore \textit{et al}. This shows that Eq.~\eqref{eq:seven} can equally describe the deflection angle near the photon sphere.

\begin{figure}
\includegraphics[width=7cm, height=.2\textheight]{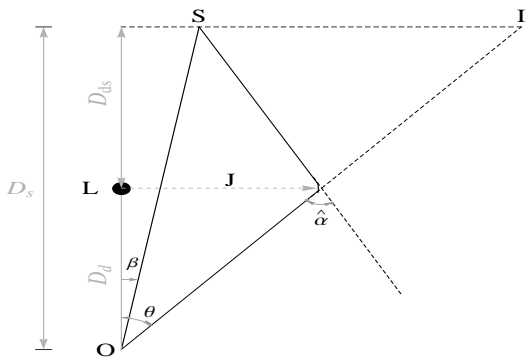}
\caption{\label{fig:lens} The lens diagram}
\end{figure}

To calculate the relativistic images we use the following lens equation ~\cite{Virbhadra2000}

\begin{equation}\label{eq:ten}
\tan \beta  = \tan \theta  - \frac{{D_{ds} }}{{D_d }}[\tan \theta  - \tan (\hat \alpha _{_D}  - \theta )]
\end{equation}
\noindent
From the lens diagram (Fig\ref{fig:lens}) we have
\begin{equation}\label{ew:nine}
\sin \theta  = \frac{J}{{D_d }}
\end{equation}

\noindent Here $\beta$ and $\theta$ be the source and the image position measured from the optic axis, $D_{ds}$ and $D_s$ be the source-lens and source-observer distance respectively. The total magnification of a circularly symmetric Gravitational lens is given by
\begin{equation}
\mu=\left(\frac{\sin{\beta}}{\sin{\theta}}
\frac{d\beta}{d\theta}\right)^{-1}
\end{equation}
and the tangential and radial magnification are given by

\begin{equation}\label{eq:eleven}
\mu_t=\left(\frac{\sin{\beta}}{\sin{\theta}}\right)^{-1} , \qquad
\mu_r=\left(\frac{d\beta}{d{\theta}}\right)^{-1}.
\end{equation}

Here we shall take an example as given in the reference ~\cite{Virbhadra2000} to calculate the relativistic images and Einstein rings etc. using the model equation ~\eqref{eq:seven}. The mass of the lens $M=2.8\times10^6 M_ \odot$ and the distance $D_d =8.5$ kpc. The lens is halfway between the point source and the observer, i.e. $D_{ds}/D_s = 1/2$.

\begin{table}
\caption {\label{tab:table2}
 Relativistic Einstein rings}
\begin{ruledtabular}
\begin{tabular}{llll}
\textrm{ring no } & $x_0$ &
$\hat{\alpha}_{_D}$ \textrm{in} $\mu$\textrm{as} & $\theta_E$ \textrm{in} $\mu$\textrm{as}\\
\colrule
\textrm{Ring I} & 1.54514737 & 2$\pi +$ 33.6963331  & 16.84816788 \\
\textrm{Ring II} & 1.50187458 & 4$\pi +$ 33.6413731  & 16.82703842 \\
\textrm{Ring III} & 1.50008023 & 6$\pi +$ 37.2240313  & 16.82699902
\end{tabular}
\end{ruledtabular}
\end{table}

\begin{figure}
\includegraphics[width=7cm, height=.2\textheight]{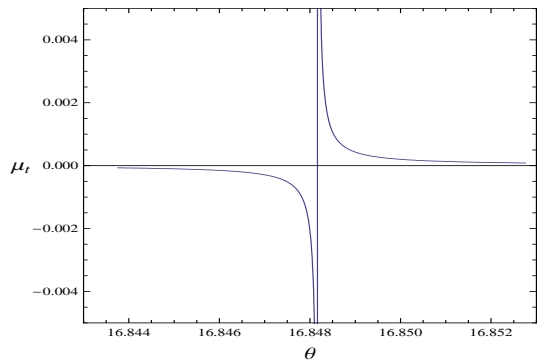}
\caption{\label{fig:tanmag} The tangential magnification near relativistic tangential critical curves. $\theta$ is in $\mu$as and $\mu_t$ is multiplied by $10^5$.}
\end{figure}

\begin{figure}
\includegraphics[width=7cm, height=.2\textheight]{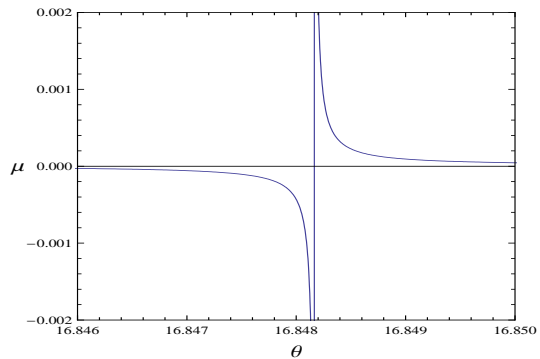}
\caption{\label{fig:totmag} The total magnification for the outermost relativistic image. $\theta$ is in $\mu$as and $\mu$ is multiplied by $10^{17}$.}
\end{figure}

The values given in Table \ref{tab:table2} shows that all the entries are approximately the same as given in Table III of Ref. ~\cite{Virbhadra2000}. We also plot the tangential magnification $\mu_t$  Fig.(~\ref{fig:tanmag}) and total magnification $\mu$ Fig.(~\ref{fig:totmag}).

\section{\label{3x} Conclusion}

The observation of relativistic images will provide a test for the success of general theory of relativity in  the strong field. We should have accurate theoretical value of deflection angle in the strong field limit to compare with observational result. We propose a model equation for the deflection angle and found that it can  correctly describe the deflection angle in the strong field limit of a Schwarzschild object . We have compared the values of deflection angle, relativistic Einstein ring, magnification etc with the values given by Bozza ~\cite{Bozza2002}, Virbhadra ~\cite{Virbhadra2000} and others. Our results almost match with those values. So we are confident that the model equation can be used to calculate different observable.


\bibliography{gravlensing}

\providecommand{\noopsort}[1]{}\providecommand{\singleletter}[1]{#1}%
\begin{thebibliography}{14}%
\makeatletter
\providecommand \@ifxundefined [1]{%
 \@ifx{#1\undefined}
}%
\providecommand \@ifnum [1]{%
 \ifnum #1\expandafter \@firstoftwo
 \else \expandafter \@secondoftwo
 \fi
}%
\providecommand \@ifx [1]{%
 \ifx #1\expandafter \@firstoftwo
 \else \expandafter \@secondoftwo
 \fi
}%
\providecommand \natexlab [1]{#1}%
\providecommand \enquote  [1]{``#1''}%
\providecommand \bibnamefont  [1]{#1}%
\providecommand \bibfnamefont [1]{#1}%
\providecommand \citenamefont [1]{#1}%
\providecommand \href@noop [0]{\@secondoftwo}%
\providecommand \href [0]{\begingroup \@sanitize@url \@href}%
\providecommand \@href[1]{\@@startlink{#1}\@@href}%
\providecommand \@@href[1]{\endgroup#1\@@endlink}%
\providecommand \@sanitize@url [0]{\catcode `\\12\catcode `\$12\catcode
  `\&12\catcode `\#12\catcode `\^12\catcode `\_12\catcode `\%12\relax}%
\providecommand \@@startlink[1]{}%
\providecommand \@@endlink[0]{}%
\providecommand \url  [0]{\begingroup\@sanitize@url \@url }%
\providecommand \@url [1]{\endgroup\@href {#1}{\urlprefix }}%
\providecommand \urlprefix  [0]{URL }%
\providecommand \Eprint [0]{\href }%
\providecommand \doibase [0]{http://dx.doi.org/}%
\providecommand \selectlanguage [0]{\@gobble}%
\providecommand \bibinfo  [0]{\@secondoftwo}%
\providecommand \bibfield  [0]{\@secondoftwo}%
\providecommand \translation [1]{[#1]}%
\providecommand \BibitemOpen [0]{}%
\providecommand \bibitemStop [0]{}%
\providecommand \bibitemNoStop [0]{.\EOS\space}%
\providecommand \EOS [0]{\spacefactor3000\relax}%
\providecommand \BibitemShut  [1]{\csname bibitem#1\endcsname}%
\let\auto@bib@innerbib\@empty
\bibitem [{\citenamefont {Darwin}(1959)}]{Darwin59}%
  \BibitemOpen
  \bibfield  {author} {\bibinfo {author} {\bibfnamefont {C.}~\bibnamefont
  {Darwin}},\ }\href@noop {} {\bibfield  {journal} {\bibinfo  {journal} {Proc.
  R. Soc. Lond. A}\ }\textbf {\bibinfo {volume} {249}},\ \bibinfo {pages} {180}
  (\bibinfo {year} {1959})}\BibitemShut {NoStop}%
\bibitem [{\citenamefont {Darwin}(1961)}]{Darwin61}%
  \BibitemOpen
  \bibfield  {author} {\bibinfo {author} {\bibfnamefont {C.}~\bibnamefont
  {Darwin}},\ }\href@noop {} {\bibfield  {journal} {\bibinfo  {journal} {Proc.
  R. Soc. Lond. A}\ }\textbf {\bibinfo {volume} {263}},\ \bibinfo {pages} {39}
  (\bibinfo {year} {1961})}\BibitemShut {NoStop}%
\bibitem [{\citenamefont {Bozza}\ \emph {et~al.}(2001)\citenamefont {Bozza},
  \citenamefont {Capozziello}, \citenamefont {Iovane},\ and\ \citenamefont
  {Scarpetta}}]{Bozza2001}%
  \BibitemOpen
  \bibfield  {author} {\bibinfo {author} {\bibfnamefont {V.}~\bibnamefont
  {Bozza}}, \bibinfo {author} {\bibfnamefont {S.}~\bibnamefont {Capozziello}},
  \bibinfo {author} {\bibfnamefont {G.}~\bibnamefont {Iovane}}, \ and\ \bibinfo
  {author} {\bibfnamefont {G.}~\bibnamefont {Scarpetta}},\ }\href@noop {}
  {\bibfield  {journal} {\bibinfo  {journal} {Gen Rel Grav}\ }\textbf {\bibinfo
  {volume} {33}},\ \bibinfo {pages} {1535} (\bibinfo {year}
  {2001})}\BibitemShut {NoStop}%
\bibitem [{\citenamefont {Bozza}(2002)}]{Bozza2002}%
  \BibitemOpen
  \bibfield  {author} {\bibinfo {author} {\bibfnamefont {V.}~\bibnamefont
  {Bozza}},\ }\href@noop {} {\bibfield  {journal} {\bibinfo  {journal} {Phy Rev
  D}\ }\textbf {\bibinfo {volume} {66}},\ \bibinfo {pages} {103001} (\bibinfo
  {year} {2002})}\BibitemShut {NoStop}%
\bibitem [{\citenamefont {Bozza}(2010)}]{Bozza2010}%
  \BibitemOpen
  \bibfield  {author} {\bibinfo {author} {\bibfnamefont {V.}~\bibnamefont
  {Bozza}},\ }\href@noop {} {\bibfield  {journal} {\bibinfo  {journal} {Gen Rel
  Grav}\ }\textbf {\bibinfo {volume} {42}},\ \bibinfo {pages} {2269} (\bibinfo
  {year} {2010})}\BibitemShut {NoStop}%
\bibitem [{\citenamefont {Amore}\ and\ \citenamefont
  {Arceo}(2006)}]{Amore2006}%
  \BibitemOpen
  \bibfield  {author} {\bibinfo {author} {\bibfnamefont {P.}~\bibnamefont
  {Amore}}\ and\ \bibinfo {author} {\bibfnamefont {S.}~\bibnamefont {Arceo}},\
  }\href@noop {} {\bibfield  {journal} {\bibinfo  {journal} {Phy Rev D}\
  }\textbf {\bibinfo {volume} {73}},\ \bibinfo {pages} {083004} (\bibinfo
  {year} {2006})}\BibitemShut {NoStop}%
\bibitem [{\citenamefont {Amore}\ and\ \citenamefont
  {Cervantes}(2007)}]{Amore2007}%
  \BibitemOpen
  \bibfield  {author} {\bibinfo {author} {\bibfnamefont {P.}~\bibnamefont
  {Amore}}\ and\ \bibinfo {author} {\bibfnamefont {M.}~\bibnamefont
  {Cervantes}},\ }\href@noop {} {\bibfield  {journal} {\bibinfo  {journal} {Phy
  Rev D}\ }\textbf {\bibinfo {volume} {75}},\ \bibinfo {pages} {083005}
  (\bibinfo {year} {2007})}\BibitemShut {NoStop}%
\bibitem [{\citenamefont {Virbhadra}\ and\ \citenamefont
  {Ellis}(2000)}]{Virbhadra2000}%
  \BibitemOpen
  \bibfield  {author} {\bibinfo {author} {\bibfnamefont {K.~S.}\ \bibnamefont
  {Virbhadra}}\ and\ \bibinfo {author} {\bibfnamefont {G.~F.~R.}\ \bibnamefont
  {Ellis}},\ }\href@noop {} {\bibfield  {journal} {\bibinfo  {journal} {Phy Rev
  D}\ }\textbf {\bibinfo {volume} {62}},\ \bibinfo {pages} {084003} (\bibinfo
  {year} {2000})}\BibitemShut {NoStop}%
\bibitem [{\citenamefont {Iyer}\ and\ \citenamefont
  {Petters}(2007)}]{Iyer2007}%
  \BibitemOpen
  \bibfield  {author} {\bibinfo {author} {\bibfnamefont {S.~V.}\ \bibnamefont
  {Iyer}}\ and\ \bibinfo {author} {\bibfnamefont {A.~O.}\ \bibnamefont
  {Petters}},\ }\href@noop {} {\bibfield  {journal} {\bibinfo  {journal} {Gen
  Rel Grav}\ }\textbf {\bibinfo {volume} {39}},\ \bibinfo {pages} {1563}
  (\bibinfo {year} {2007})}\BibitemShut {NoStop}%
\bibitem [{VLB()}]{VLBI}%
  \BibitemOpen
  \href@noop {} {\enquote {\bibinfo {title} {Constellation-{X} web page:
  constellation.gsfc.nasa.gov; maxim web page: maxim.gsfc.nasa.gov},}\
  }\BibitemShut {NoStop}%
\bibitem [{\citenamefont {Ulvestad}(1999)}]{Ulvestad1999}%
  \BibitemOpen
  \bibfield  {author} {\bibinfo {author} {\bibfnamefont {J.~S.}\ \bibnamefont
  {Ulvestad}},\ }\href@noop {} {\bibfield  {journal} {\bibinfo  {journal} {New
  Astron Rev}\ }\textbf {\bibinfo {volume} {43}},\ \bibinfo {pages} {531}
  (\bibinfo {year} {1999})}\BibitemShut {NoStop}%
\bibitem [{\citenamefont {Weinberg}(1972)}]{Weinberg}%
  \BibitemOpen
  \bibfield  {author} {\bibinfo {author} {\bibfnamefont {S.}~\bibnamefont
  {Weinberg}},\ }\href@noop {} {\emph {\bibinfo {title} {Gravitation and
  Cosmology}}}\ (\bibinfo  {publisher} {Wiley},\ \bibinfo {year}
  {1972})\BibitemShut {NoStop}%
\bibitem [{\citenamefont {Bisnovatya-Kogan}\ and\ \citenamefont
  {Yu}(2008)}]{Bisnovatya}%
  \BibitemOpen
  \bibfield  {author} {\bibinfo {author} {\bibfnamefont {G.~S.}\ \bibnamefont
  {Bisnovatya-Kogan}}\ and\ \bibinfo {author} {\bibfnamefont {T.~O.}\
  \bibnamefont {Yu}},\ }\href@noop {} {\bibfield  {journal} {\bibinfo
  {journal} {APJ}\ }\textbf {\bibinfo {volume} {51}},\ \bibinfo {pages} {99}
  (\bibinfo {year} {2008})}\BibitemShut {NoStop}%
\bibitem [{\citenamefont {Misner}\ \emph {et~al.}(1973)\citenamefont {Misner},
  \citenamefont {Thorne},\ and\ \citenamefont {Wheeler}}]{Misner}%
  \BibitemOpen
  \bibfield  {author} {\bibinfo {author} {\bibfnamefont {C.~W.}\ \bibnamefont
  {Misner}}, \bibinfo {author} {\bibfnamefont {K.~S.}\ \bibnamefont {Thorne}},
  \ and\ \bibinfo {author} {\bibfnamefont {J.~A.}\ \bibnamefont {Wheeler}},\
  }\href@noop {} {\emph {\bibinfo {title} {Gravitation}}}\ (\bibinfo
  {publisher} {W A Freeman and Company},\ \bibinfo {year} {1973})\BibitemShut
  {NoStop}%
\end{thebibliography}%

\end{document}